\documentclass[aps,prd, 10pt, twocolumn,nofootinbib,showpacs]{revtex4}
\usepackage[usenames,dvipsnames]{xcolor} 
\usepackage{amsmath,amssymb}

\begin{document}

\title{The generalized second law for fermionic and bosonic fields}
\author{Koray D\"{u}zta\c{s}}
\email{koray.duztas@okan.edu.tr}
\affiliation{Faculty of Engineering and Natural Sciences, \.{I}stanbul Okan University, 34959 Tuzla \.{I}stanbul, T\"{u}rkiye}
\begin{abstract}
We evaluate the validity of the generalied second law for Kerr black holes perturbed by fermionic and bosonic fields.  We derive that the critical frequency for a test field below which  the area of a Kerr black hole would decrease, coincides with the superradiance limit which pertains to bosonic fields. The fact that the absorption of fermionic fields with arbitrarily low frequencies is allowed, leads to a generic violation of the generalized second law as both the black hole and the environment lose entropy. The result does not contradict the proof of the area theorem which pre-assumes the validity of the null energy condition. We also construct a thought experiment involving bosonic fields to check whether the minimum increase in the area can compensate for the decrease in the entropy of the environment. We minimize the entropy increase by considering a black hole at the extremal limit, perturbed by a bosonic field at the superradiance limit. We show that the generalized second law remains valid for bosonic fields that satisfy the null energy condition. The result does not require the employment of entropy bounds when one assigns von Neumann entropy to test fields.
\end{abstract}
\pacs{04.70.Dy}
\maketitle
\section{Introduction}
The proof of the area theorem dates back the early 1970s which is undisputedly the most revolutionary era in black hole physics. Hawking proved that if two black holes merge to form a single one, the area of the final black hole  is larger than the sum of the initial areas provided that the null energy condition is satisfied and there are no naked singularities in the exterior region \cite{hawkarea}. 
\begin{equation}
\delta A \geq 0.
\label{hawkarea1}
\end{equation}
The result also applies to  a single black hole perturbed by a test particle or field. This is strongly reminiscent of the second law of thermodynamics which asserts that the entropy of a closed system cannot decrease. The striking similarity between the area theorem and the second law inspired Bekenstein to propose that a black hole must have an entropy proportional to its area, and the total entropy of the black hole and the environment system should not decrease \cite{beken1,beken2,beken3}
\begin{equation}
\Delta (S_{\rm{BH}}+S_{\rm{ext}})\geq 0,
\label{beken1}
\end{equation}
where $S_{\rm{BH}}$ and $S_{\rm{ext}}$ represent the entropy of the black hole (or Bekenstein-Hawking entropy)  and the entropy of the exterior region, respectively. The relation (\ref{beken1}) is referred to as the generalized second law (GSL) of black hole thermodynamics.

Following Bekenstein's proposal Bardeen, Carter and Hawking formulated the laws of black hole mechanics by establishing analogies between the surface gravity and the temperature, and between the area and entropy \cite{thermo}. However, they emphasized that these analogies should not be mistaken for a physical correspondence as they assumed that the  effective temperature of a black hole should be absolute zero. Maintaining his scepticism of Bekenstein's proposal, Hawking embarked on a study of particle creation during the formation of a black hole. He considered quantum fields in the time-dependent phase of a gravitationally collapsing body. His stunning discovery that black holes emit radiation dramatically altered the perception of black holes and their interaction with the environment \cite{hawkingorig}. Hawking derived that  the average number of particles reaching the future null infinity per unit time can be expressed as a function of the frequency $\omega$ and the angular momentum numbers $l,m$
\begin{equation}
N_{\omega lm}=\frac{\Gamma_{lm}(\omega)}{\exp [2\pi (\omega - m\Omega)/\kappa] \mp 1},
\label{hawk1}
\end{equation}
where $\Omega$ is the angular velocity of the event horizon and $\kappa$ is the surface gravity. The minus and plus signs apply to bosons and fermions respectively.  $\Gamma_{lm}$ is the  probability that a wave in exactly the same mode $(\omega,l,m)$ would be absorbed by the black hole, which is also known as the grey-body factor. (See \cite{sakalli} for a review).  The radiation is thermal and provides a precise value for the temperature and the proportionality between the surface area and entropy which was yet undetermined
\begin{equation}
T=\frac{\kappa}{2\pi}, \quad S=\frac{A(kc^3)}{4(G \hbar)}.
\end{equation}

An astonishing feature of the entropy-area proportionality is that it forges a link between statistical mechanics, gravitation and quantum mechanics as it involves Boltzmann's constant $k$, Newton's gravitational constant $G$ and Planck's constant $\hbar$. The precise definitions of temperature and entropy have elevated black hole thermodynamics from a conceptual analogy to a rigorously established physical theory. 

Hawking radiation is incompatible with the area theorem (\ref{hawkarea1}) as it leads the black hole to shrink and lose entropy. However, GSL is not violated as radiation increases the entropy of the exterior region. One can prove the validity of GSL for Hawking radiation evaluating the  von Neumann entropy for the thermal density matrix of each mode (see e.g. \cite{pagerev})
\begin{equation}
\delta S_{\rm{rad}}=-\sum P_n \ln P_n =(N \pm 1) \ln (1 \pm N) - N\ln N,
\label{vonneuman1}
\end{equation}
where $P_n$ is the probability of $n$ particles being in the mode $(\omega,l,m)$, and $\pm$ signs refer to bosonic and fermionic radiation, respectively. The decrease in the entropy of the black hole is compensated by the increase in the entropy of the exterior region in the process of Hawking radiation; thus GSL remains valid. Bekenstein constructed an alternative thought experiment where a spherical test body of mass/energy $(\delta M)$ and radius $R$ is lowered sufficiently near a black hole, then dropped. The absorption of the test body leads to an increase in the surface area of the black hole, the minimum value of which is given by
\begin{equation}
\Delta A_{\rm{min}}=8\pi (\delta M)R.
\end{equation}
The increase in the entropy of the black hole should compensate for the decrease in the entropy of the exterior region to assure that GSL is not violated. To avoid the violation of GSL, Bekenstein conjectured the existence of an upper bound for the entropy of a test body \cite{bekenbound}
\begin{equation}
S\leq 2\pi (\delta M)R.
\label{bekenbound1}
\end{equation}
The upper bound (\ref{bekenbound1}) re-assures the validity of GSL in Bekenstein's thought experiment. However, the existence of a universal upper bound has been subject to controversy \cite{pagecomment,deutch}. Unruh and Wald  argued that Bekenstein bound is unnecessary as the buoyancy effect of Hawking radiation could prevent one from lowering a test body sufficiently close to black hole so that its energy can be extracted \cite{waldunruhbuo1,waldunruhbuo2}. However, the buoyancy effect of Hawking radiation would be negligible in a process of lowering a macroscopic object. Bekenstein gave rebuttals against Unruh and Wald \cite{bekenbuo1,bekenbuo2}. On the other hand, to lower a macroscopic object quasi-statically near a black hole and to keep it stationary, work must be done by an external agent which would lead the entropy of the system to increase. Incorporating the increase in the entropy during the interaction, GSL remains valid without referring to Bekenstein bound \cite{gaowald,kanzi}. We would like to note that whether it is necessary to invoke Bekenstein's upper bound to prove the validity GSL is not essential in addressing the fundamental question whether such a universal entropy bound exists.

A covariant generalization of Bekenstein's bound was proposed by Bousso \cite{bousso1}. If $A$ is the area of a closed surface, the entropy on the hypersurface generated by null geodesics orthogonal to $A$ with non-negative expansion satisfies $S\leq A/4$. The requirement that the geodesics should be non-expanding is the covariant generalization of the intuitive notion of ``inside'' for a closed surface. Though proofs were given for the covariant entropy bound \cite{covproof1,covproof2}, they were criticized for their assumptions on the entropy current \cite{wallten}. Despite the lack of rigorous proofs, the entropy bounds and GSL are indispensable tools to probe the semi-classical and quantum features of black holes  \cite{gsl1,gsl2,gsl3,gsl4,gsl5,gsl6,gsl7,gsl8,gsl9}. In particular, the combination of  entropy bounds with unitary evolution in quantum mechanics  directly implies the holographic principle, thus opens the gateway for the realm of quantum gravity \cite{boussorev}.

Here, we evaluate the validity of the area theorem and GSL for Kerr black holes interacting with massless fermionic and bosonic fields. The main difference between the two types of fields is that bosonic fields satisfy the null energy condition, while fermionic fields do not. As a consequence, there exists a lower limit for the frequency of bosonic fields to allow their absorption by the black hole. For the modes with a lower frequency, the absorption probability of bosonic fields becomes negative; i.e. the fields are reflected back to infinity with a larger amplitude. This is the well-known effect of superradiance, which only occurs for bosonic fields. The result is derived by evaluating the reflection and transmission coefficients for wave equations in Kerr-like spacetimes \cite{super1,super2,super3,super4,super5}. For fermionic fields, the absorption probability remains positive for arbitrarily low frequencies. In that respect, the course of the interaction with fermionic fields is fundamentally different. We evaluate the validity of the area theorem and GSL for Kerr black holes perturbed by fermionic fields in section (\ref{fermi}).

In attempts to violate GSL, one usually checks if the increase in the entropy of a black hole can compensate for the decrease in the entropy of the exterior region. For that purpose, Hod constructed a thought experiment where a nearly extremal Reissner-Nordstr\"{o}m black hole is perturbed by a test field \cite{hodgsl}. He derived that the increase in the entropy of the black hole approaches zero in the extremal limit, which cannot compensate for the decrease in the entropy of the exterior region even if Bekenstein's entropy bound is valid. To remedy the violation of GSL, he employed a form of Hoop conjecture. In section (\ref{boson}), we construct a parallel thought experiment where a nearly extremal Kerr black hole is perturbed by a bosonic test field. We incorporate the second order variations into the analysis for nearly extremal black holes, since  the event horizon may appear to be destroyed when one ignores their contribution. We assign von Neumann entropy to the fields and check whether GSL remains valid in the extremal limit. We finally interpret our results and discuss possible resolutions.

\section{The area theorem and GSL for fermionic fields}
\label{fermi} 
Hawking's proof of the area theorem (\ref{hawkarea1}) pre-assumes the validity of the null energy condition. However, fermionic fields do not satisfy the null energy condition. From this point, one cannot directly infer that fermionic fields violate the area theorem. This merely renders us inconclusive about the application of the area theorem to fermionic fields. In this section, we check the validity of the area theorem and GSL for Kerr black holes perturbed by fermionic fields. 

In his seminal work Teukolsky showed that the wave equations  can be separated in Kerr background for each spin $(s=0,s=1/2,s=1,s=3/2,s=2)$ in the form \cite{teuk}
\begin{equation}
\Psi=e^{-i\omega t}e^{im\phi}S(\theta)R(r).
\end{equation}
The contributions of the test fields in this form, to the mass and angular momentum parameters of the Kerr space-time are related by
\begin{equation}
\delta J=\frac{m}{\omega}\delta M.
\label{deltajdeltam}
\end{equation}
The area of a Kerr black hole is
\begin{equation}
A=4\pi(r_+^2 +a^2)=8\pi (M^2+\sqrt{M^4-J^2}),
\end{equation}
where $r_+=M+\sqrt{M^2-a^2}$ is the radial coordinate of the event horizon, and $a=J/M$ is the Kerr parameter. Throughout this work, we adopt natural units where $G=\hbar=c=1$. In this system the angular momentum parameter $J$ and the surface area $A$ have dimensions of $M^2$; the event horizon radius $r_+$ and the Kerr parameter $a=J/M$ have dimensions of $M$; and the frequency $\omega$, the angular velocity $\Omega$, and the surface gravity $\kappa$ carry dimensions of $1/M$. We envisage that a test field with frequency $\omega$ and azimuthal number $m$ is incident on a Kerr black hole. The contributions of the test field to  the mass and angular momentum parameters of the space-time are related by (\ref{deltajdeltam}).  We do not impose a priori whether the test field is fermionic or bosonic; nor whether it satisfies the null energy condition.  Instead, we examine whether the interaction can result in a negative variation of the black hole area  ($\delta A<0)$. 
\begin{equation}
\frac{\delta A}{8 \pi}= \left(2M +\frac{2M^2}{\sqrt{M^2 - \frac{J^2}{M^2} }} \right) \delta M + \left(\frac{-J}{M\sqrt{M^2 - \frac{J^2}{M^2}}} \right)\delta J .
\label{deltaA1}
\end{equation}
We substitute $\delta J=(m/\omega)\delta M$ and $a=J/M$ in (\ref{deltaA1})
\begin{equation}
\delta A=8 \pi \frac{\left[2M (M+\sqrt{M^2-a^2}) -a\left(\frac{m}{\omega}\right)\right] }{\sqrt{M^2-a^2}} \delta M ,
\label{delatA2}
\end{equation}
which implies
\begin{eqnarray}
&& \omega \geq \frac{ma}{2Mr_+}=m\Omega \Rightarrow  \delta A \geq 0, \nonumber \\
&&\omega<\frac{ma}{2Mr_+}=m\Omega \Rightarrow \delta A <0, 
\label{deltaAomega}
\end{eqnarray}
where $\Omega$ is the angular velocity of the event horizon. If the frequency of a test field is below the critical value derived in (\ref{deltaAomega}), the area of the black hole will decrease as a result of the interaction. The question at this stage is whether a test field in this mode will be absorbed by the black hole. The critical value derived in (\ref{deltaAomega}) identically coincides with the limiting frequency for superradiance $\omega_{\rm{sl}}=m\Omega$, which occurs for bosonic fields. If the frequency of a bosonic test field is below the superradiance limit, the field is reflected back to infinity with a larger amplitude. No net absorption of these modes occurs. However, if a fermionic field is incident on a black hole, the absorption probability remains positive even for arbitrarily low frequencies. The general expression for the absorption probabilities for bosonic and fermionic fields scattering from Kerr black holes were derived by Page in his seminal work \cite{page}. In a recent study on Hawking radiation at the zero temperature limit, we have derived the absorption probabilities for bosonic and fermionic fields and evaluated them in the extremal limit \cite{hawkzero}. For a scalar field with $(s=0,l=1)$, the absorption probability takes the form
\begin{equation}
\Gamma^{\rm{B}}_{(s=0,l=1)}=\frac{1}{18}\left( \frac{A\omega}{2\pi} \right)^3(\omega -m\Omega) \left[ \kappa^2 + (\omega -m\Omega)^2 \right],
\label{probboson}
\end{equation}
where $\kappa$ is the surface gravity and the superscript (B) refers to bosonic fields.  The factor $(A\omega)^3$ has dimensions of $M^3$; $(\omega -m\Omega)$ has dimensions of $1/M$; and $\kappa^2$ and $(\omega -m\Omega)^2$ have dimensions of $1/M^2$. One can verify that the absorption probability $\Gamma$ and the particle number $N$ are dimensionless parameters. The absorption probability (\ref{probboson}) becomes negative for $(\omega < m\Omega)$. These modes are not absorbed by the black hole. By means of (\ref{deltaAomega}), $\delta A \geq 0$ for bosonic fields which satisfy the null energy condition. The result is in accord with Hawking's area theorem, which apply to the perturbations satisfying the null energy condition. 

For a fermionic field with $s=1/2$, one derives \cite{hawkzero}
\begin{equation}
\Gamma^{\rm{F}}_{(l=s=1/2)}=\frac{1}{4} \left( \frac{A \omega}{2\pi} \right)^2 [\kappa^2 +4(\omega - m\Omega)^2].
\label{probls1half}
\end{equation}
The expression (\ref{probls1half}) is positive definite for all $\omega$. The absorption of the modes with $\omega < m\Omega$ is allowed for fermionic fields, which leads the area of the black hole to decrease.  We would like to note that Hawking's proof of the area theorem pre-assumes the validity of the null energy condition, which is not satisfied by fermionic fields. Therefore the possibility that $\delta A$ can be negative for fermionic fields does not contradict the proof of the area theorem. 

In the thought experiments to challenge the validity of GSL, one usually checks if the increase in the black hole entropy is sufficiently large to compensate for the decrease of the entropy of the exterior region. However, if a fermionic field with $\omega < m\Omega$ is absorbed by the black hole, both the black hole and the environment lose entropy. The entropy lost by the environment can be calculated using (\ref{vonneuman1}) for fermionic fields. Note that $\Gamma$ remains positive for $\omega < m\Omega$, and $(0<N<\Gamma <1)$
\begin{equation}
\delta S^{\rm{F}}=(N-1)\ln(1-N) -N\ln(N) \geq 0.
\label{entropyfermi}
\end{equation}
When a fermionic field with frequency $\omega < m\Omega$ is absorbed by the black hole, the environment experiences a net loss of entropy. (Otherwise we could have concluded that the environment effectively gains entropy.) In this interaction, both the black hole and its surroundings lose entropy. Although we assign von Neumann entropy to the test fields, this assignment is not crucial to establishing the entropy loss of the environment. The absorption of fermionic fields with $\omega < m\Omega$ thus results in a general violation of the generalized second law, as it leads to a simultaneous decrease in entropy for both the black hole and the external region.
\section{Extremal limit and Bosonic fields}
\label{boson} 
We alluded to Bekenstein's thought experiment, where one lowers a test body adiabatically near a black hole to challenge GSL. The possible violation of GSL is fixed by Bekenstein's entropy bound (\ref{bekenbound1}). Relatively recently Hod constructed a thought experiment where one lowers a test body adiabatically towards a nearly extremal black hole. He claimed that the increase in the entropy would be drastically small in the extremal limit, and GSL would be violated even if the entropy bound is valid \cite{hodgsl}. He employed a form of the Hoop conjecture to re-assure the validity of GSL. In this section we construct a similar thought experiment where one perturbs a nearly extremal Kerr black hole with a scalar test field satisfying the null energy condition. We check whether the minimal increase in the entropy of the black hole in the extremal limit can compensate for the entropy decrease in the exterior region. Note that  the result (\ref{deltaAomega})  implies $\delta A=0$ for $\omega=m\Omega$. We minimize the increase in the entropy by lowering the frequency of the test field arbitrarily close to the superradiance limit. 

We start with a nearly extremal Kerr black hole \\ parametrized as
\begin{equation}
M^2-\frac{J^2}{M^2}=M^2 \epsilon^2,
\label{param1}
\end{equation}
where the dimensionless parameter $\epsilon \ll 1$ parametrizes the closeness to extremality. We perturb the black hole with a test field which contributes to the mass and angular momentum parameters as follows:
\begin{equation}
\delta M=M\eta, \quad \delta J=\frac{m}{\omega}\delta M.
\label{paramfield1}
\end{equation}
To justify the test field approximation, we impose that $\eta \ll 1$. Note that we use different small parameters for the closeness of the black hole to extremality, and the energy carried by the test field. In the extremal limit only $\epsilon$ approaches zero, but not $\eta$.

Note that (\ref{param1}) implies $J^2=M^4(1-\epsilon^2)$. Using 
\[
(1-\epsilon^2)^{1/2}\simeq (1-\frac{\epsilon^2}{2}),
\]
one derives
\begin{equation}
J=Ma=M^2(1-\frac{\epsilon^2}{2}).
\label{param1a}
\end{equation}
Note that the parametrization (\ref{param1}) directly implies $r_+=M(1+\epsilon)$. Using the expression for $a$ given in (\ref{param1a}), the angular velocity of the event horizon of a nearly extremal black hole can be calculated as
\begin{equation}
\Omega=\frac{a}{2Mr_+}=\frac{1-\epsilon^2/2}{2M(1+\epsilon)}.
\label{angvelepsilon}
\end{equation}
We assume the null energy condition is satisfied. To minimize the increase in the area, we consider a test field with frequency
\begin{equation}
\omega=\frac{m}{2M(1+\epsilon)}>m\Omega.
\label{omegafield}
\end{equation}
For bosonic fields satisfying the null energy equation, we have to require $\omega > m\Omega$ to assure that the field is absorbed by the black hole. For a test field with frequency given in (\ref{omegafield}) the contribution to the angular momentum parameter is given by
\begin{equation}
\delta J=\frac{m}{\omega}\delta M=2M^2(\eta +\eta\epsilon +O(3)).
\label{testfield1}
\end{equation}
The increase in the area of the black hole is
\begin{equation}
\Delta A=A_{\rm{fin}}-A_{\rm{in}},
\end{equation}
where 
\begin{equation}
A_{\rm{in}}=8\pi \left( M^2 + M\sqrt{M^2-\frac{J^2}{M^2}} \right)=8\pi M^2(1+\epsilon),
\label{areainitial}
\end{equation} 
and where we used the parametrization (\ref{param1}). After the interaction, the final area is
\begin{equation}
A_{\rm{fin}}=8\pi \left( M_{\rm{fin}}^2 + M_{\rm{fin}}\sqrt{M_{\rm{fin}}^2-\frac{J_{\rm{fin}}^2}{M_{\rm{fin}}^2}} \right),
\label{areafinal1}
\end{equation} 
where 
\begin{eqnarray}
M_{\rm{fin}} &=& M+\delta M +\delta^2 M, \nonumber \\
M_{\rm{fin}}^2&=&M^2+ (\delta M)^2 +2M\delta M +2M \delta^2 M, \nonumber \\
J_{\rm{fin}} &=& J+ \delta J+ \delta^2 J, \nonumber \\
J_{\rm{fin}}^2&=&J^2+ (\delta J)^2 +2J\delta J +2J \delta^2 J,
\label{mfinjfin}
\end{eqnarray}
where $M_{\rm{fin}}^2$ and  $J_{\rm{fin}}^2$ are expanded to second order. Previously, we derived that the first order variations due to test fields with frequencies near the superradiance limit lead to over-spinning of Kerr black holes, which is equivalent to the expression in square roots in (\ref{areafinal1}) being negative \cite{overspin}. However, the problem is fixed when one incorporates the contribution of the second order variations \cite{spin2}. Therefore we have to consider the contribution of the second order variations in our analysis. We are going to proceed by evaluating (\ref{areafinal1}). Note that to second order $J_{\rm{fin}}^2 / M_{\rm{fin}}^2$ is equal to
\begin{equation}
\frac{J_{\rm{fin}}^2}{M_{\rm{fin}}^2}=\frac{J^2+ (\delta J)^2 +2J\delta J +2J \delta^2 J}{M^2}\left( 1-2\eta +3\eta^2 -2\frac{\delta^2 M}{M} \right),
\label{jfinmfinsquare}
\end{equation}
where we used (\ref{mfinjfin}) and substituted $\delta M=M\eta$. We substitute the expression for $\delta J$ from (\ref{testfield1}) and proceed. To second order one derives
\begin{eqnarray}
M_{\rm{fin}}^2 -\frac{J_{\rm{fin}}^2}{M_{\rm{fin}}^2}&=&M^2\epsilon^2 +2M^2\eta^2 -4M^2\eta \epsilon  \nonumber \\
&+&4M\delta^2 M - 2\delta^2 J.
\label{mfinjfin2}
\end{eqnarray}
Note that the result (\ref{mfinjfin2}) would be negative for $\eta \sim \epsilon$, if we had ignored the contribution of the second order variations. In \cite{spin2}, we evaluated the Sorce-Wald condition \cite{sorcewald} for second variations of a Kerr black hole and derived that
\begin{equation}
2M\delta^2 M-\delta^2 J =\frac{(\delta J)^2}{2M^2}.
\end{equation}
Substituting the expression for $\delta J$ from (\ref{testfield1})
\begin{equation}
4M\delta^2 M-2\delta^2 J =4M^2\eta^2 +O(3),
\end{equation}
substituting this to (\ref{mfinjfin2})
\begin{equation}
M_{\rm{fin}}^2 -\frac{J_{\rm{fin}}^2}{M_{\rm{fin}}^2}=M^2\epsilon^2 +6M^2\eta^2 -4M^2\eta \epsilon .
\end{equation}
We can evaluate $\Delta A$ in the extremal limit $\epsilon \to 0$
\begin{equation}
\lim_{\epsilon \to 0}\Delta A=8\pi M^2 \left[(2+\sqrt{6})\eta +(1+\sqrt{6})\eta^2+ \frac{2}{M}\delta^2 M \right].
\label{deltaAresult}
\end{equation}
We observe that $\delta A$ is first order in $\eta$, in the extremal limit $\epsilon \to 0$. The increase in the area/entropy of the black hole does not vanish in the extremal limit, as a consequence of the fact that the energy of the perturbations at the asymptotically flat infinity does not depend on the black hole parameters.  

To derive the von Neumann entropy for the test field, note that
\begin{equation}
\kappa=\frac{r_+-r_-}{2(r_+^2 + a^2)}=\frac{2M\epsilon}{4M^2(1+\epsilon)}=\frac{\epsilon -\epsilon^2 +O(3)}{2M}.
\label{kappaepsilon}
\end{equation}
Also for the frequency (\ref{omegafield}) assigned to the test field
\begin{equation}
\omega - m\Omega=\frac{m\epsilon^2}{4M(1+\epsilon)}=\frac{m\epsilon^2}{4M} + O(3),
\label{omegaminus}
\end{equation}
which is derived by substracting the expression for $m\Omega$ in (\ref{angvelepsilon}) from (\ref{omegafield}). Note that the azimuthal number $m$ can be $\{0,1\}$ for $l=1$. Since the modes with $m=0$ do not contribute to the angular momentum parameter, we proceed with $m=1$. This leads to
\begin{equation}
\exp [2\pi (\omega - m\Omega)/\kappa] - 1=\exp [\pi \epsilon] -1 =\pi \epsilon + O(2).
\label{exponentialepsi}
\end{equation}
We evaluate the absorption probability $\Gamma$ for bosonic fields derived in (\ref{probboson}). The initial parameters of the black hole determine the absorption probability as the test field is incident on the event horizon. Note that
\[ \left( \frac{A\omega}{2\pi}\right)^3=8M^3 ,
\]
using the expressions for $A$  in (\ref{areainitial}); and for $\omega$ in (\ref{omegafield}). This leads to
\begin{equation}
\Gamma^{\rm{B}}_{(s=0,l=1)}=\frac{1}{18} (8M^3) \frac{\epsilon^2}{4M} \left[ \frac{\epsilon^2}{4M^2} + \frac{\epsilon^4}{16M^2} \right],
\label{probboson1}
\end{equation}
where we have used (\ref{kappaepsilon}) and (\ref{omegaminus}). Note that the leading term in $\Gamma^{\rm{B}}$ is fourth order in $\epsilon$. We can evaluate $N$ in the limit $\epsilon \to 0$ for bosonic test fields with frequency given in (\ref{omegafield})
\begin{equation}
N=\frac{\Gamma}{\pi \epsilon + O(2)}\sim \frac{\epsilon^4}{\epsilon}\sim \epsilon^3, \quad \lim_{\epsilon \to 0}N=0,
\label{limitN}
\end{equation}
where we have used (\ref{exponentialepsi}) and (\ref{probboson1}). As $N \to 0$, the von Neumann entropy for a bosonic test field approaches zero
\begin{equation}
\delta S_{\rm{rad}}=(N + 1) \ln (1 + N) - N\ln N, \quad \lim_{N \to 0} \delta S_{\rm{rad}}=0 .
\label{entropyfield}
\end{equation}
In the extremal limit, the von Neumann entropy associated with test fields of frequency $\omega \gtrsim m\Omega$ tends toward zero while the increase in the black hole’s entropy remains first order in $\eta$. This supports the conclusion that the generalized second law holds without the need to invoke an entropy bound. Crucially, the increase in the area of the black hole does not vanish in the extremal limit because we consider the energy of the test fields as measured at infinity when determining their contribution to the mass parameter of the  black hole.
\section{Conclusions}
In this work we constructed thought experiments to test the validity of GSL, when a Kerr black hole is perturbed by fermionic and bosonic fields. We first derived the critical value for the frequency of a test field below which $\delta A$ can be negative. This value identically coincides with the superradiance limit $\omega_{\rm{sl}}=m\Omega$, which occurs for bosonic fields. However, fermionic fields with frequencies below the critical value are absorbed by the black hole. The absorption of fermionic fields with frequency below the superradiance limit leads the area of the black hole to decrease; i.e. the black hole loses entropy. Simultaneously, the environment loses entropy as the test field is absorbed. The process of the absorption of fermionic fields with $\omega < m\Omega$ identifies a generic violation of GSL, unlike the previous attempts where the increase in the area of the black hole fails to compensate for the decrease in the entropy of the environment. To avoid any confusion we would like to note that the result $\delta A<0$ for fermionic fields does not contradict Hawking's area theorem, since fermionic fields do not satisfy the null energy condition which is required to prove $\delta A>0$.

For the cases satisfying the null energy condition, the attempts to violate GSL are based on minimizing the increase in the area so that it cannot compensate for the decrease in the entropy of the exterior region. The most common way to remedy the possible violation of GSL in this type of thought experiments is to invoke the entropy bounds first proposed by Bekenstein \cite{bekenbound}. In a relatively recent work, Hod claimed that the increase in the area converges to zero in the extremal limit. Consequently the violation of GSL cannot be fixed by Bekenstein's entropy bound in this case \cite{hodgsl}. We constructed a similar thought experiment involving bosonic test fields. We minimized the increase in the area by imposing that the black hole is nearly extremal, and the frequency of the test field is close to the superradiance limit. We incorporated the second order variations, since the event horizon appears to be destroyed when one ignores their contribution in these limits. We evaluated GSL in the limit where the black hole approaches extremality. We did not encounter the pathologies mentioned by Hod. The increase in the area does not approach zero in the extremal limit. GSL remains valid without the need to invoke the entropy bounds, if one assigns von Neumann entropy to test fields. The main difference from the previous work is that we assert that the mass/energy of the perturbations at the asymptotically flat infinity  contributes to the mass parameter of the space-time. Hod considers the red-shifted energy of a test body at the horizon which depends on $(r_+ -r_-)$, and approaches zero at the extremal limit.  However, the energy of a test body at infinity should be independent of the black hole parameters. The vanishing of the area/entropy increase at the extremal limit is an artefact due to incorporating the red-shifted energy at the horizon.

The generalized second law (GSL) is widely regarded as a reliable principle in black hole thermodynamics. Its foundation in statistical and quantum mechanics has even led to proposals suggesting it could replace traditional energy condition assumptions. For instance, Wall demonstrated a singularity theorem by substituting energy conditions with GSL \cite{wallsing}. However, the analysis presented in this work indicates that the validity of GSL is not entirely independent of energy conditions within the framework of current black hole physics and its semi-classical extensions. Specifically, GSL holds true only when the null energy condition is satisfied. In earlier work, we showed that the absorption of fermionic fields with $\omega < m\Omega$ results in a generic violation of the cosmic censorship conjecture, as $M_{\rm{fin}}^2 < J_{\rm{fin}}$ \cite{generic,spinhalf,threehalves}. This process also violates the third law of black hole thermodynamics, allowing the black hole to be driven continuously to extremality and beyond. These findings suggest that fermionic fields, in their current treatment, may be fundamentally incompatible with black hole physics. It is expected that a complete quantum theory of gravity will reconcile fermionic field behaviour with black hole dynamics, thereby restoring the validity of GSL, the third law, and cosmic censorship.

\end{document}